# A new and improved ephemeris for the hot subdwarf eclipsing binary SDSS J082053.53+000843.4

D. Pulley, D. Smith, G. Faillace, A. Watkins

**Abstract**

SDSS J082053.53+000843.4 belongs to the HW Vir family of short period binary systems and was first identified by Geier in 2011. Whilst three subsequent papers have focused on the morphology of this system, little has been published on the system period and its constancy. Here we provide the first published times of minima together with a revised ephemeris and a binary period, 0.0962404(1)d, that is four orders of magnitude more precise than previous declared values.

## What is SDSS J082053.53+000843.4?

SDSS J082053.53+000843.4, for brevity here referred to as J08205, is a $15^{th}$ magnitude short period, 0.0962404(1)d (~$2^h$ $18.6^m$), detached eclipsing binary system thought to consist of a post main sequence subdwarf B star with a brown dwarf companion. J08205 belongs to a subclass of the HW Vir family of very short period (< 4 hours) eclipsing binary systems comprising a hot subdwarf O or B type star and a dwarf M star or brown dwarf companion. The hot subdwarf is a core helium burning star with a thin hydrogen shell and is situated at the extreme end of the horizontal branch on the H-R diagram.

Formation of an HW Vir type system has been speculatively explained as the evolution of a binary system consisting of two main-sequence stars of different masses. As soon as the more massive star reaches the red-giant phase and fills its Roche lobe, mass transfer takes place to the companion star. If the mass transfer becomes unstable, the envelope of the more massive star will engulf the companion star and form a common envelope. Drag forces on the two stellar objects transfers angular momentum from the binary pair to the common envelope. When the envelope has gained sufficient energy it is ejected and the momentum loss from the binary system results in a much reduced binary separation and period. The end product is a close system containing the core of the giant, which becomes a hot sdB star, and a cooler main-sequence or brown dwarf companion. The 2013 summary paper by Zorotovic et al.[1], and references contained therein, describe the common envelope process and lists thirteen other HW Vir type systems. Some of these systems have shown variations in their eclipse timings leading a number of investigators to speculate that these period variations are the result of light travel time effects caused by an orbiting substellar object(s), see for example Qian et al.[2] The formation and survival of circumbinary substellar objects prior to or post the common envelope phase is an area of ongoing research. When quasi-cyclical variations in period have been observed magnetic quadrupole effects, as described by Applegate,[3] have been put forward as an alternative explanation. However it is considered by most investigators that the available energy budget to support magnetic effects is too small to explain the observed variations.

**Historical Observations**

J08205 was first identified by Geier et al.[4] in 2011 as part of the Massive Unseen Companions to Hot Faint Underluminous Stars from SDSS (MUCHFUSS) project. Geier, Schaffenroth et al. published further findings in 2011.[5][6][7]

Spectroscopic and light curve analysis of J08205 presented in these papers predicts that the binary companion is a brown dwarf with a mass of approximately $0.068 M_\odot$.

The effective temperatures of the primary and secondary components compute to 26,700K and ~2,500K respectively. The separation of the components is given as

$0.7R_\odot$, equivalent to approximately 1.5 Earth - Moon separations. The spectroscopic analysis is not well constrained for the secondary component and the figures quoted above assume a canonical mass of the primary sdB of $0.47M_\odot$. Geier derives a system ephemeris of...[5]

$$T_{HJD} = 2455147.8564(6) + 0.096(1)*E \qquad (1)$$

where the ephemeris is given as Heliocentric Julian Date (HJD) and E is the Epoch. There are no known published eclipse timings for J08205 and thus no indication of whether light travel time effects are apparent in this system. This short paper publishes the first comprehensive set of eclipse timings for J08205 and, if transit time variations are found, will lay the foundation for exploring the presence of a potential third body in a system where magnetic quadrupole effects generated by the brown dwarf are expected to be much reduced or non-existent.

**New Observations and Analysis**

The November 2009 ephemeris of Geier (Equation 1) could not predict with sufficient accuracy the times of minima for the 2014/15 observing season so, on the 16[th] and 20[th] December 2014, we made two preliminary observing runs. Having identified a new reference time of minimum we then captured 19 primary minima and 6 secondary minima on 11 nights between the 25[th] December 2014 and 22[nd] February 2015. Our observations in the latter part of February 2015 provided complete light curves for this system in both B and V filters.

Images were calibrated with flat, darks and bias fields or, as appropriate, LCOGT pipeline; photometry was performed using MaxIm DL. Reference stars were selected from APASS Catalogue (AAVSO) to have a close as possible B-V colour and magnitude match to the very blue target.

The data points of a typical light curve for J08205 are shown in Figure 1 together with the calculated light curve, shown in red, using the PHEOBE V0.31a software package.[8]   The flat bottom of both the primary and secondary minimum confirm that the eclipse is total and the rising and falling shoulders between the eclipses confirm the strong irradiation effects of the very hot sdB primary on the nearby and much cooler brown dwarf secondary.  These results are consistent with Geier et al. who reported similar total primary and secondary eclipses[5]

**Computing an Improved Ephemeris**

We used the Kwee and van Woerden procedure [9] contained in the software suite Peranso, and verified using Minima software package,[10] to determine the times of minima from each of our observing nights.  These calculated minima and their uncertainties are listed in Table 1 where the times have been converted to Baricentric Julian Dates (BJD).   Unweighted linear regression analysis was performed on the 19 primary minima to determine a much improved ephemeris over that published by Geier [5]...

$$T_{BJD} = 2457016.61118\ (5) + 0.0962404(1)*E \qquad (2)$$

The period calculated in our new ephemeris, Equation 2, has an uncertainty of less than 10ms and provides four orders of magnitude improvement in precision over the last published estimate (see in Equation 1) of 0.096(1)d.  Using our new ephemeris we calculated the O - C (Observed minus Calculated) residuals but, as can be seen from Figure 2, no trend over this short time span can be observed.

## Conclusions and Future Work

Our observations of J08205 provide a revised and improved binary period of 0.0962404(1)d which reduces the uncertainty by four orders of magnitude over previous estimates. At present there is insufficient data to confirm whether this period is stable, or subject to some form of long term variability. It is our intention to monitor J08205 over time in order to establish whether this system exhibits period variations often seen in HW Vir type systems and, if such variations exist, to identify their source.

## Acknowledgements

This work makes use of observations from the LCOGT network of telescopes[11] and of the APASS database maintained on the AAVSO website.

## References

1       Zorotovic, M., et al. A&A 549, A95, 2013

2       Qian, S-B., MNRAS 436, 1408-1414, 2013

3       Applegate, J., ApJ, 385, 621-629, 1992 February 1

4       Geier et al. AIP Conf. Proc. Vol 1331, 2011 (arXiv:1012.3839)

5       Geier, S.,et al., ApJ Letters, Vol 731 No. 2, 2011 (arXiv:1103.1989)

6       Geier, S., et al. 2012 ASP Conf. Ser., 452 153. (arXiv:1112:2929)

7       Schaffenroth, V., et al. AIP Conf. Ser., 1331, 2011. (arXiv:1102.0502)

8       Prsa, A and Zwitter, T, 2005, ApJ 628 426

9       Kwee K. K. & van Woerden, H., 1956 Bull. Astron. Instit. of Netherlands, 12, 327

10      http://members.shaw.ca/bob.nelson/software1.htm

11    Brown, T. M. et al., Publications of the Astronomical Society of the Pacific, 2013, Volume 125, issue 931, pp.1031-1055 "Las Cumbres Observatory Global Telescope Network",

**Address:** c/o British Astronomical Association, Burlington House, Piccadilly, London W1J 0DU

**DP:** david@davidpulley.co.uk;  **DS:** astro@beechwoodhouse.me.uk;  **AM:** americowatkins@aol.com;
**GF:** binarygfaillace@aol.com

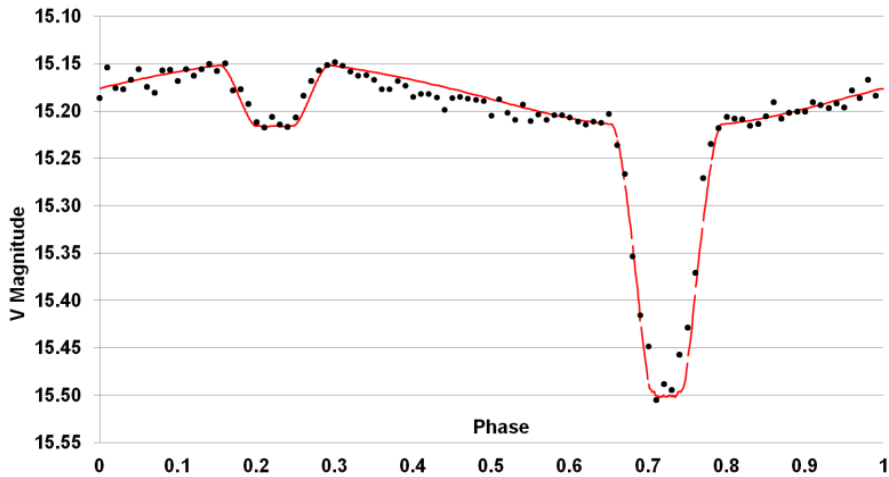

Fig. 1. Phased V-band light curve (black circles) of J082053.53+000843.4 taken on 20150222 with the 1m telescope at the Sutherland Obsevatory, S.A. The red curve is the system modelled with PHEOBE

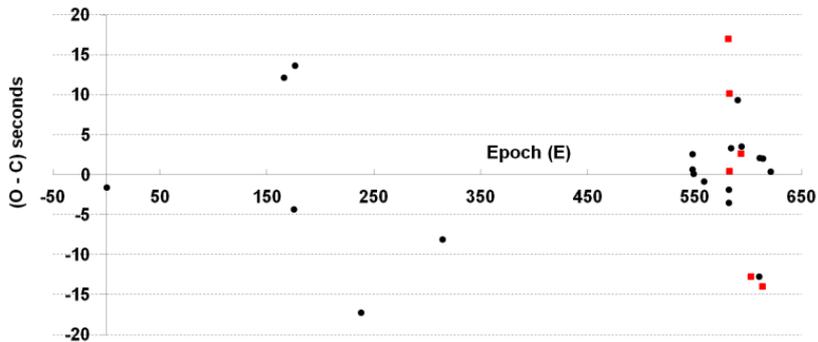

Fig. 2. (O - C) residuals calculated from the new ephemeris of Eq. 2. This ephemeris was calculated from the primary minima and shown as he black markers. The secondary minima are shown as red squares.

| Date | MJD (+2400000) | MBJD (+2400000) | E | Minima Type | Uncertainty (days) | Filter | Observatory |
|---|---|---|---|---|---|---|---|
| 25/12/2014 | 57016.605970 | 57016.611163 | 0 | I | 0.000093 | Clear | 2 |
| 10/01/2015 | 57032.581368 | 57032.587238 | 166 | I | 0.000079 | V | 1 |
| 10/01/2015 | 57033.447316 | 57033.453211 | 175 | I | 0.000350 | V | 2 |
| 11/01/2015 | 57033.543762 | 57033.549659 | 176 | I | 0.000231 | V | 2 |
| 17/01/2015 | 57039.510170 | 57039.516209 | 238 | I | 0.000108 | Clear | 2 |
| 24/01/2015 | 57046.824455 | 57046.830590 | 314 | I | 0.000081 | V | 3 |
| 15/02/2015 | 57069.345109 | 57069.350978 | 548 | I | 0.000081 | V | 5 |
| 15/02/2015 | 57069.345088 | 57069.350957 | 548 | I | 0.000028 | B Bessell | 5 |
| 15/02/2015 | 57069.441324 | 57069.447190 | 549 | I | 0.000202 | V | 5 |
| 16/02/2055 | 57070.403748 | 57070.409584 | 559 | I | 0.000091 | B Bessell | 5 |
| 19/02/2015 | 57072.569437 | 57072.575200 | 581.5 | II | 0.000119 | V | 4 |
| 19/02/2015 | 57072.617341 | 57072.623103 | 582 | I | 0.000053 | V | 4 |
| 19/02/2015 | 57072.617322 | 57072.623084 | 582 | I | 0.000046 | V | 4 |
| 19/02/2015 | 57072.665602 | 57072.671362 | 582.5 | II | 0.000099 | V | 4 |
| 19/02/2015 | 57072.665489 | 57072.671249 | 582.5 | II | 0.000213 | V | 4 |
| 19/02/2015 | 57072.809889 | 57072.815643 | 584 | I | 0.000115 | B Bessell | 6 |
| 19/02/2015 | 57073.387421 | 57073.393156 | 590 | I | 0.000129 | B Bessell | 5 |
| 20/02/2015 | 57073.724198 | 57073.729919 | 593.5 | II | 0.000191 | B Bessell | 6 |
| 20/02/2015 | 57073.772330 | 57073.778050 | 594 | I | 0.000056 | B Bessell | 6 |
| 21/02/2015 | 57074.590216 | 57074.595905 | 602.5 | II | 0.000052 | V | 4 |
| 21/02/2015 | 57075.312047 | 57075.317709 | 610 | I | 0.000125 | V | 5 |
| 21/02/2015 | 57075.408464 | 57075.414122 | 611 | I | 0.000074 | V | 5 |
| 22/02/2015 | 57075.648889 | 57075.654537 | 613.5 | II | 0.000103 | V | 4 |
| 22/02/2015 | 57075.697195 | 57075.702842 | 614 | I | 0.000035 | V | 4 |
| 22/02/2015 | 57076.370887 | 57076.376506 | 621 | I | 0.000100 | V | 5 |

Table 1. Calculated times of minima of J082053.53+000843.4 derived from our observations taken between 2014/12/25 and 2015/02/22

Observatories: (1) 0.32m f/8 T18, iTelescopes Nerpio, Spain (2) 0.36m f/7.7, Astrognosis Observatory, UK (3) 0.61m f/10 Sierra Stars Observatory, Carson Valley, California (4) 1m Las Cumbres Observatory, Cerro Tololo, Chile (5) 1m Sutherland, S.A. (6) 1m MacDonald, USA